\documentclass[aps,prl,reprint,superscriptaddress,showkeys]{revtex4-1}
\usepackage{amsmath}
\usepackage{graphicx}
\usepackage{natbib}

\begin{document}

\title{Anisotropy-assisted non-scattering coherent absorption of surface plasmon-polaritons}

\author{Anton I. Ignatov}
\affiliation{Moscow Institute of Physics and Technology, 9 Institutskiy per., Dolgoprudny 141700, Russia}
\affiliation{All-Russia Research Institute of Automatics, 22 Sushchevskaya, Moscow 127055, Russia}

\author{Igor A. Nechepurenko}
\affiliation{Moscow Institute of Physics and Technology, 9 Institutskiy per., Dolgoprudny 141700, Russia}
\affiliation{All-Russia Research Institute of Automatics, 22 Sushchevskaya, Moscow 127055, Russia}

\author{Denis G. Baranov}
\email[]{denis.baranov@phystech.edu}
\affiliation{Moscow Institute of Physics and Technology, 9 Institutskiy per., Dolgoprudny 141700, Russia}
\affiliation{All-Russia Research Institute of Automatics, 22 Sushchevskaya, Moscow 127055, Russia}

\date{\today}


\begin{abstract}
The ability to control propagation of electromagnetic guided modes lies at the heart of integrated nanophotonics. Surface plasmon-polaritons are a class of guided modes which can be employed in integrated optical systems.
Here, we present a theoretical design of a coherent surface plasmon absorber which can perfectly harvest energy of coherently incident surface plasmons without parasitic scattering into free space modes. Excitation of free space modes which usually accompanies scattering of a surface plasmon by an interface boundary is avoided due to specially tailored anisotropy of the absorber.  The concept of coherent SPP absorber is analyzed numerically for spatially non-uniform and finite-size structures. We believe that our results will be important for the development of integrated nanoplasmonic systems.
\end{abstract}

\maketitle

\section{Introduction}
Surface plasmon polaritons (SPPs) are localized electromagnetic excitations which can propagate along a metal-dielectric interface. Due to their localized nature, they find a great variety of fascinating applications including subwavelength guiding, imaging, sensing and light harvesting~\cite{Polo2011,Zayats2005}. SPPs are also a promising candidate for information transfer between components of an all-optical chip. In order to effectively manipulate propagating SPPs within an integrated nanophotonic structure specific control elements are needed which allow one to guide, reflect, bend, and focus SPP waves as well as to perform logical operations. There has been a significant theoretical and experimental success towards developing such nanostructures in the recent years~\cite{Yin2005,Lopez-Tejeira2007,Gramotnev2010,Shin2006,Liu2009,Brongersma,Tanaka, Fu}. One more essential element to handle SPPs in the context of data processing is the device capable of coherent perfect absorption (CPA) of propagating SPPs, which can serve as a modulator or an optical switch in a plasmonic circuit~\cite{Bozhevolnyi}.

The notion of CPA was firstly introduced in Ref.~\cite{Chong}. A CPA is an electromagnetic system in which perfect absorption of radiation is mediated via \emph{interference of two incident waves}~\cite{Longhi}.
In the plane wave configuration the CPA can be implemented as a symmetric slab of lossy medium coherently illuminated at both sides (input ports). When parameters of the system are tuned to the CPA condition and the two incident light waves have proper phase difference, the reflected and transmitted waves interfere destructively and the incoming energy is perfectly absorbed inside the slab. The concept of CPA allows to control light with another beam of light via interference and therefore enables realization of \emph{linear} optical switches and logical gates~\cite{Zheludev12}.
The experimental demonstration of CPA was reported in Refs.~\cite{Wan,Zanotto} for Fabry-Perot cavities and in Ref.~\cite{Yoon} for a plasmonic guided mode system.
Coherent absorption was also investigated for single nanoparticles, what required coherent $4\pi$ illumination of the particle with a specific mode of electromagnetic field~\cite{Sentenac, Noh}, and for point dipole sources~\cite{Klimov12, Klimov13}.

One intuitive way to extend the idea of CPA to the SPP configuration is to integrate an interface between two isotropic lossy materials supporting propagation of SPPs into the external plasmonic waveguide what at certain length of the absorbing interface should result in the destructive interference of the two incident SPPs. However, such a naive approach would not work, since incidence of an SPP on a boundary between two different metal-insulator (MI) interfaces is always accompanied by scattering into the continuous spectrum of free space modes~\cite{Stegeman1981,Stegeman1983}, what is strongly unwanted in integrated optical systems.
A few approaches for achieving perfect absorption in the waveguide configuration have been theoretically explored recently. They rely either on incorporation of resonant nanoantennas within or close to the waveguide~\cite{Park2015,Bruck}, or on dissipative couplers~\cite{Zanotto2015, Grote}. Still, scattering of incident energy into the free space modes has not been questioned in those works.

Here we demonstrate theoretically that scattering-free coherent SPP absorber may be realized as an interface between two uniaxial media with specific permittivity tensors. Parasitic scattering of SPPs into free space modes is eliminated due to anisotropy of the absorbing layer, leading to incident plasmons being reflected and refracted at the boundary of two interfaces in a manner similar to plane waves. Under coherent illumination of the anisotropic plasmonic waveguide from both sides CPA regime is realized leading to complete absorption of the incident plasmons. The concept of SPP CPA is verified numerically via full-wave simulations for the case of finite size anisotropic absorbing metamaterials. Our findings provide a new approach for realization of a plasmon coherent absorber and will facilitate the development of scattering-free plasmon components for integration in all-optical circuits.

\section{Theory}
Certain approaches for minimization of parasitic SPP scattering by an abrupt MI interface change have been suggested. Those include use of layered structures made of dielectric isotropic materials~\cite{Bezus2014}, matched anisotropic metamaterials~\cite{Elser2008}, and magnetic ($\mu \ne1$) materials~\cite{Novitsky2010}. Only the latter two methods enable exact suppression of parasitic scattering. Since magnetic materials are not available in optics, we will incorporate findings of Ref.~\cite{Elser2008} in order to extend the CPA concept to the plasmonic configuration. 

Figure 1 presents a schematic structure of the SPP CPA. It is formed by an interface between two anisotropic media, supporting an SPP mode, integrated into a common MI plasmonic waveguide. Absorption of incident SPPs occurs inside the central lossy region of the structure. The interface between the two anisotropic materials is geometrically adjusted to those of the external waveguides.

\begin{figure}[!t]
\includegraphics[width=1\columnwidth]{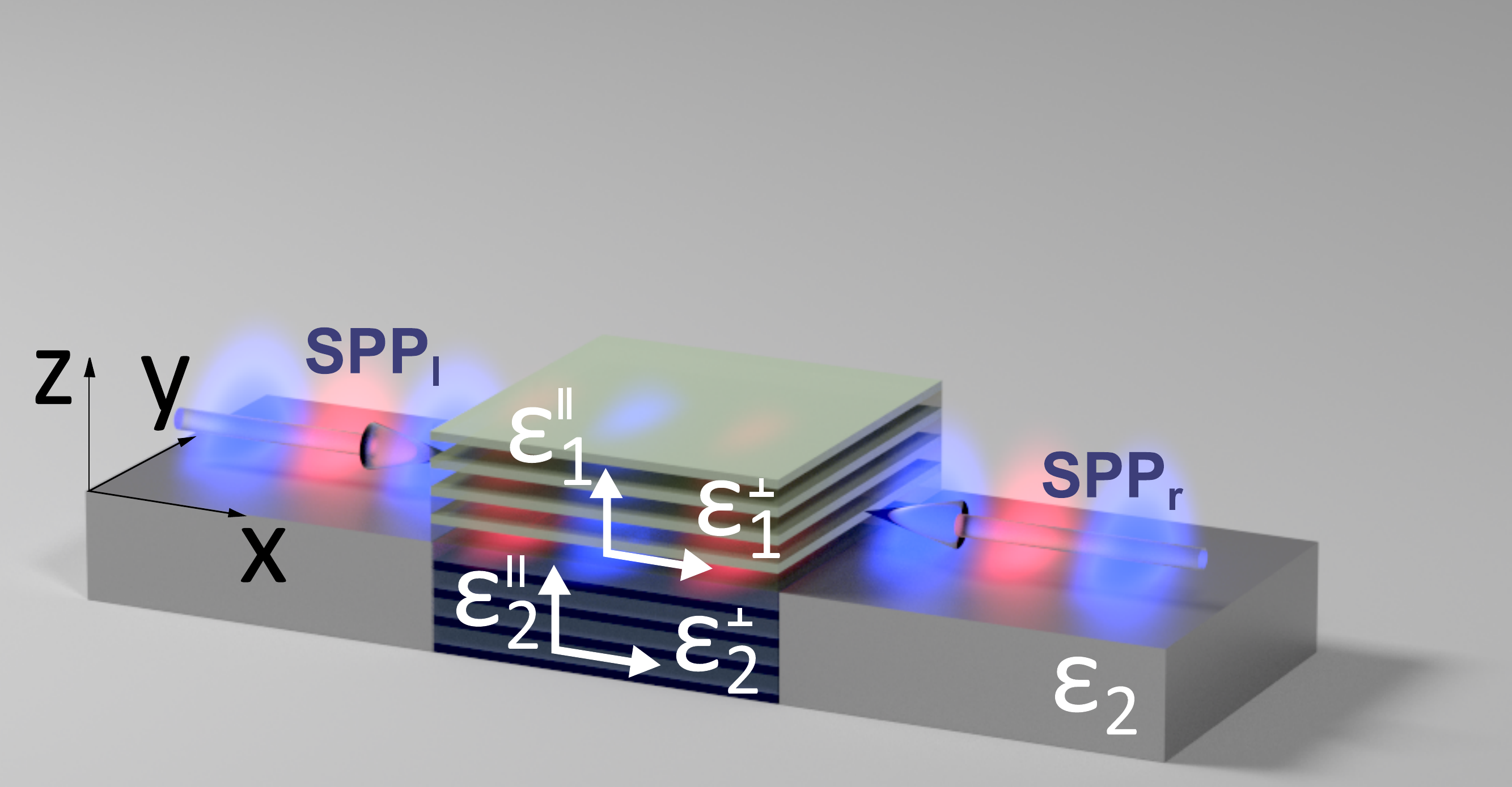}
\caption{A schematic of the proposed SPP CPA. Two identical metal-dielectric interfaces form the external region, from which two coherent surface plasmons (SPP$_r$ and SPP$_l$) impinge on an anisotropic absorber, formed by layered metal-dielectric metamaterials. Incident SPPs form a standing wave inside the structure, and the energy is completely absorbed without being scattered.}
\label{fig1}
\end{figure}

We begin our analysis with recalling the conditions upon which incidence of a TM-polarized SPP on a boundary between two interfaces is not accompanied by parasitic scattering. SPP propagating in the external waveguide is characterized by its wave vector ${\mathbf{k}} = \left( {{k_x},0,{k_z}} \right)$, whose longitudinal and normal components are given by the well-known expressions:
\begin{equation}
{k_x} = \frac{\omega }
{c}\sqrt {\frac{{{\varepsilon _1}{\varepsilon _2}}}
{{{\varepsilon _1} + {\varepsilon _2}}}},~ k_z^{(1,2)} = \sqrt {{\varepsilon _{1,2}}\frac{{{\omega ^2}}}
{{{c^2}}} - k_x^2} ,
\label{eq1}
\end{equation}
where 1 and 2 denote the upper and lower half-spaces, respectively. Wave vector ${\bf{\kappa}} = \left( {{\kappa_x},0,{\kappa_z}} \right)$ of an SPP propagating at the interface between two uniaxial materials is described by a pair of more complicated expressions:
\begin{equation}
{\kappa _x} = \frac{\omega }
{c}\sqrt {\frac{{\varepsilon _1^\parallel \varepsilon _2^\parallel \left( {\varepsilon _1^ \bot  - \varepsilon _2^ \bot } \right)}}
{{\varepsilon _1^ \bot \varepsilon _1^\parallel  - \varepsilon _2^ \bot \varepsilon _2^\parallel }}} ,~\kappa _z^{(1,2)} = \sqrt {\varepsilon _{1,2}^ \bot \left( {\frac{{{\omega ^2}}}
{{{c^2}}} - \frac{{\kappa _x^2}}
{{\varepsilon _{1,2}^\parallel }}} \right)}.
\label{eq2}
\end{equation}
\begin{figure*}[!t]
\includegraphics[width=2\columnwidth]{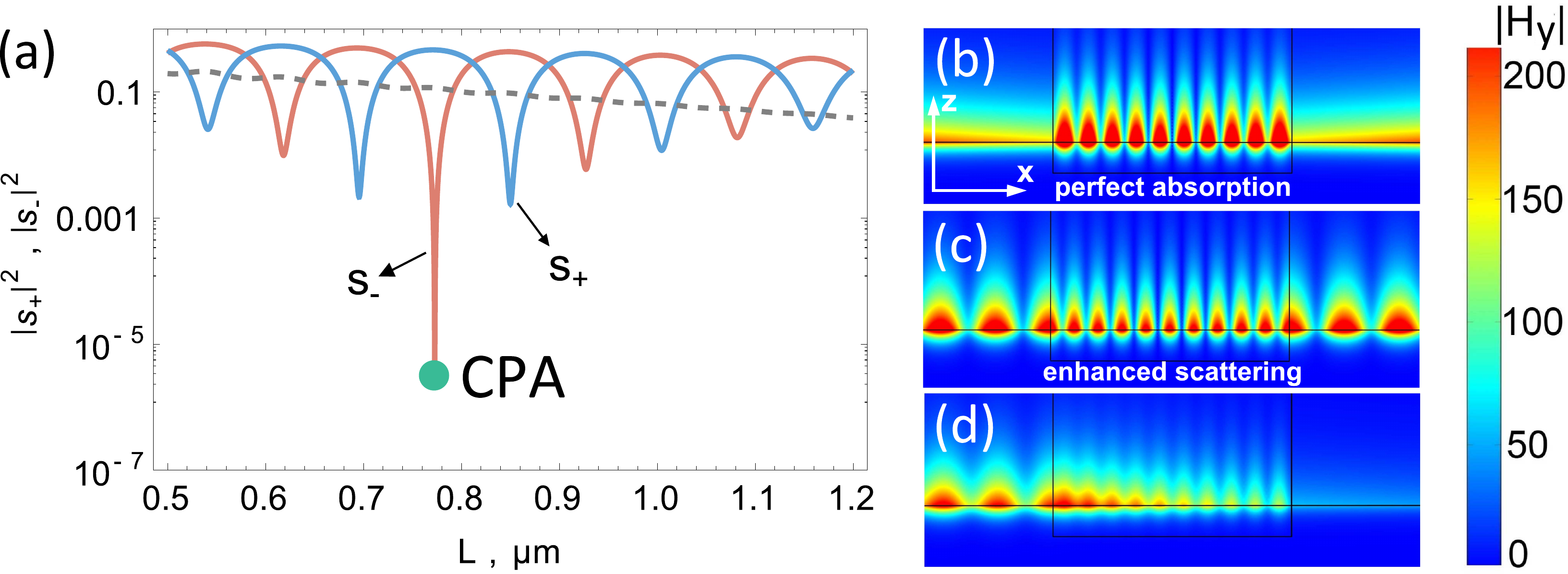}
\caption{(a) Absolute values of the symmetric ($s_+$) and antisymmetric ($s_-$) $S-$matrix eigenvalues of the anisotropic scattering-free plasmonic interface vs length of the interface. Sharp dip of the antisymmetric mode eigenvalue indicates the occurrence of the perfect absorption. Dashed line shows the transmission for a single port illumination. (b) Map of the magnetic field magnitude $|H_y|$ for the CPA solution at the slab length $L= 770$~nm. (c) The corresponding magnetic field map for the symmetric incidence. (d) The corresponding picture when only one side of the system (left) is illuminated. The color bar to the right is in arbitrary units.}
\label{fig2}
\end{figure*}

Here, $\varepsilon _{1,2}^ \bot $ and $\varepsilon _{1,2}^\parallel $ stand for the in- and out-of-plane permittivity components of two anisotropic materials (normal and parallel to the optical axis of the material). Scattering into free space modes is avoided upon SPP incidence if spatial behavior of SPP field in $z-$ direction is identical in both regions, i.e., $k_z^{(i)} = \kappa _z^{(i)}$. In Ref.~\cite{Elser2008} it was shown that this requirement reduces to the following equalities:
\begin{equation}
  {\varepsilon _1} = \varepsilon _1^ \bot ,~~
  {\varepsilon _2} = \varepsilon _2^ \bot ,~~
  \frac{{\varepsilon _1^\parallel }} {{{\varepsilon _1}}} = \frac{{\varepsilon _2^\parallel }} {{{\varepsilon _2}}}.
\label{eq3}
\end{equation}

Given that the matching conditions~(\ref{eq3}) are satisfied, amplitudes of the reflected and transmitted plasmons (with respect to transverse $y-$component of the magnetic field) under normal incidence are given by the expressions analogous to the Fresnel formulas for the case of plane wave incidence:
\begin{equation}
r = \frac{{{{{k_x}} \mathord{\left/
 {\vphantom {{{k_x}} {{\varepsilon _2}}}} \right.
 \kern-\nulldelimiterspace} {{\varepsilon _2}}} - {{{\kappa _x}} \mathord{\left/
 {\vphantom {{{\kappa _x}} {\varepsilon _2^\parallel }}} \right.
 \kern-\nulldelimiterspace} {\varepsilon _2^\parallel }}}}
{{{{{k_x}} \mathord{\left/
 {\vphantom {{{k_x}} {{\varepsilon _2}}}} \right.
 \kern-\nulldelimiterspace} {{\varepsilon _2}}} + {{{\kappa _x}} \mathord{\left/
 {\vphantom {{{\kappa _x}} {\varepsilon _2^\parallel }}} \right.
 \kern-\nulldelimiterspace} {\varepsilon _2^\parallel }}}},~
t = \frac{{2{{{k_x}} \mathord{\left/
 {\vphantom {{{k_x}} {{\varepsilon _2}}}} \right.
 \kern-\nulldelimiterspace} {{\varepsilon _2}}}}}
{{{{{k_x}} \mathord{\left/
 {\vphantom {{{k_x}} {{\varepsilon _2}}}} \right.
 \kern-\nulldelimiterspace} {{\varepsilon _2}}} + {{{\kappa _x}} \mathord{\left/
 {\vphantom {{{\kappa _x}} {\varepsilon _2^\parallel }}} \right.
 \kern-\nulldelimiterspace} {\varepsilon _2^\parallel }}}}.
\label{eq5}
\end{equation}
Correspondingly, amplitudes of the plasmons reflected and transmitted through a non-scattering interface between two anisotropic materials of finite length $L$ may be expressed as
\begin{equation}
R = \frac{{r\left( {1 - \exp \left( {2i{\kappa _x}L} \right)} \right)}}
{{1 - {r^2}\exp \left( {2i{\kappa _x}L} \right)}},~T = \frac{{\exp \left( {i{\kappa _x}L} \right)\left( {1 - {r^2}} \right)}}
{{1 - {r^2}\exp \left( {2i{\kappa _x}L} \right)}},
\label{eq6}
\end{equation}
which is analogous to usual expression for plane wave reflection and transmission amplitudes with wavenumber $\kappa_x$ inside a slab.


When the system is irradiated from both sides, amplitudes of scattered and incident SPPs on the left (subscript $l$) and right (subscript $r$) sides of the system are related via the scattering matrix:
\begin{equation}
\left( {\begin{array}{*{20}{c}}
   {s_r^{\rm out}}  \\
   {s_l^{\rm out}}  \\
 \end{array} } \right) = \hat S\left( {\begin{array}{*{20}{c}}
   {s_r^{\rm in}}  \\
   {s_l^{\rm in}}  \\
 \end{array} } \right),
 \label{eq7}
\end{equation}
where the $S-$matrix of the two-port system is defined according to $\hat S = \left( {\begin{array}{*{20}{c}}
   R & T  \\
   T & R  \\
 \end{array} } \right)$. In this picture, coherent perfect absorption, i.e., absence of scattered plasmons ($s_{r,l}^{\rm out}=0$), is defined as a zero of the scattering matrix eigenvalue occurring on the real frequencies axis~\cite{Chong,Baldacci2015}:
\begin{equation}
s_\pm=R \pm T=0.
\end{equation}
Here, signs '+' and '-' correspond to the symmetric and antisymmetric eigenmodes of the slab, respectively, having 0 or $\pi$ phase difference between the left and right channels. Hereafter we will focus on the antisymmetric eigenmode, while treatment of the symmetric solution is analogous. The problem is then to find the system parameters, including permittivity tensors components and the slab length, which yield zero of the the $s_-$ eigenvalue.

\section{Results}
In order to demonstrate our approach, we now apply it to a specific SPP structure. For the MI waveguide, from which two input SPPs are incident, we choose permittivities $\varepsilon_1=1$ and $\varepsilon_2=-4.5+0.92i$, respectively, corresponding to vacuum and Ag permittivity at $\lambda=400$~nm. All calculations below are performed for this wavelength. This results in SPP wavelength $\lambda_{\rm SPP}=355$~nm with the propagation length of $\approx 2$~$\mu$m. With this choice of MI interface, we find that matching conditions~(\ref{eq3}) are exactly satisfied for the following permittivity tensors of the integrated anisotropic structure: $\varepsilon _{1}^ \bot =1$, $\varepsilon _{1}^ \parallel = 5.28 + 0.016i$, $\varepsilon _{2}^ \bot = -4.5 + 0.92i$, $\varepsilon _{2}^ \parallel = -23.8 + 4.9i$. 

Figure~\ref{fig2}(a) shows the system eigenvalues ${\left| {{s_ \pm }} \right|^2}$ as a function of the absorbing slab length $L$. Dashed line in Fig.~\ref{fig2}(a) shows the transmission amplitude $|T|^2$ for a single-port illumination. Sharp dip experienced by the antisymmetric eigenvalue at $L\approx 770$~nm clearly indicates occurrence of the CPA condition. Remarkably, at the same time the orthogonal symmetric mode eigenvalue is of the order of 1. In other words, change of the relative phase difference between two incident SPPs transforms the system from the perfectly absorbing state (magnetic field distribution of the CPA mode is presented in Fig.~\ref{fig2}(b)) to strongly scattering state, magnetic field distribution for which is shown in Fig.~\ref{fig2}(c). The magnetic field distributions were obtained by numerical modeling in COMSOL. Such behavior was previously highlighted for other CPA systems~\cite{Chong, Wan}.

The corresponding magnetic field distribution created by a single-port illumination is presented in Fig.~\ref{fig2}(d), where substantial fields of the transmitted SPP can be seen on the opposite side of the absorber. Spatial modulation of the incident field along the $x-$axis also indicates strong reflected signal. This behavior points to crucial importance of the destructive interference and coherent illumination for achieving the total absorption.

Strong anisotropy required for the CPA regime is usually not available in natural media. However, multilayer metamaterials provide a powerful tool for achieving a broad range of effective permittivity values in the in- and out-of-plane directions. The effective permittivities of anisotropic metamaterials are related to layers parameters by the widely used mixing formulas~\cite{Milton}:
\begin{equation}
\begin{gathered}
  \varepsilon _j^ \bot  = {f_j}\varepsilon _j^{(m)} + \left( {1 - {f_j}} \right)\varepsilon _j^{(d)}, \hfill \\
  \varepsilon _j^ \parallel  = \frac{1}
{{{{{f_j}} \mathord{\left/ {\vphantom {{{f_j}} {\varepsilon _j^{(m)}}}} \right. \kern-\nulldelimiterspace} {\varepsilon _j^{(m)}}} + {{\left( {1 - {f_j}} \right)} \mathord{\left/
 {\vphantom {{\left( {1 - {f_j}} \right)} {\varepsilon _j^{(d)}}}} \right. \kern-\nulldelimiterspace} {\varepsilon _j^{(d)}}}}}, \hfill \\ 
\end{gathered} 
\label{eqeff}
\end{equation}
where $\varepsilon _{1,2}^{(m)}$ and $\varepsilon _{1,2}^{(d)}$ are the permittivities of the metallic and dielectric layers in the first and second metamaterial, and $f_{1,2}$ is the respective fraction of the metallic component. These formulas result in the desired effective permittivities of the anisotropic metamaterials with the following set of layers parameters: 
$\varepsilon _{1}^{(m)}=-2.94 + 0.01i$, $\varepsilon _{1}^{(d)}=2.53 + 0.002i$, $f_1=0.28$, 
$\varepsilon _{2}^{(d)}=10.06 + 3.06i$, $\varepsilon _{2}^{(m)}=-10.54 + 0.026i$, $f_2=0.71$. The above values for metallic components $\varepsilon _{1,2}^{(m)}$ are achievable in the visible with noble metals and aluminum. The permittivity value of $2.53 + 0.002i$ is easily attainable with some polymers, while the value of $10.06 + 3.06i$ is close to that of amorphous carbon at $\sim 400$~nm~\cite{carbon}.

\begin{figure}[!t]
\includegraphics[width=.8\columnwidth]{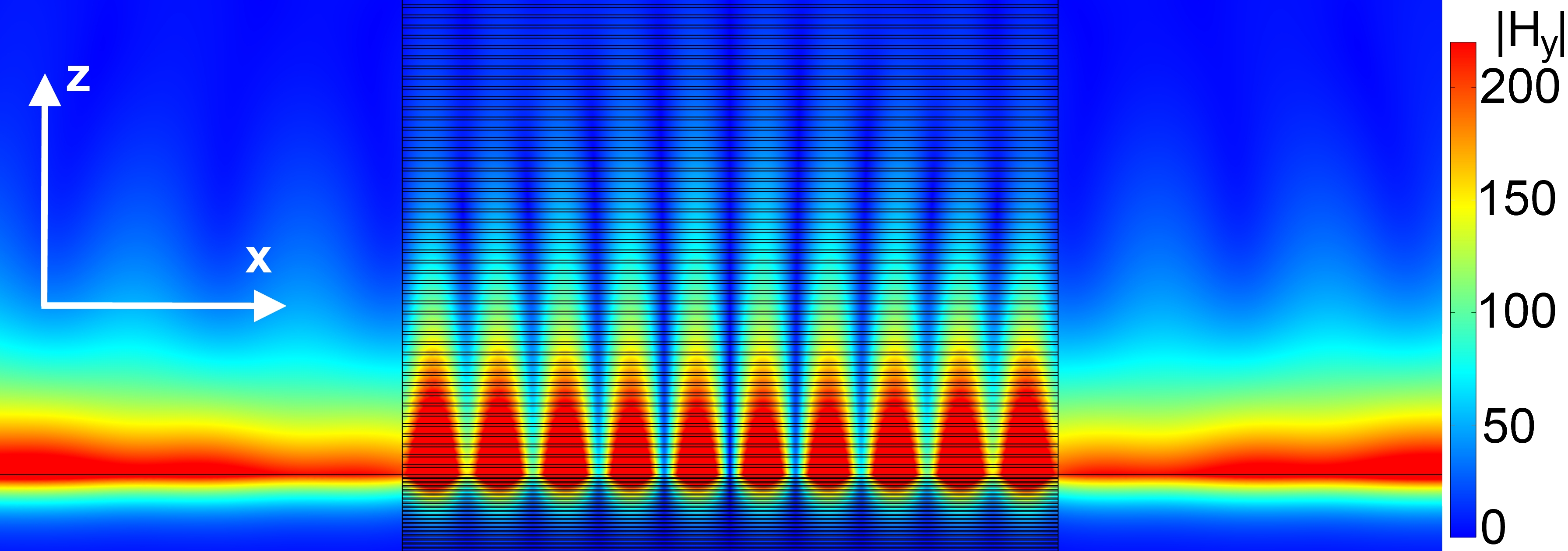}
\caption{Scattering of the CPA mode by a layered finite-periodicity system. Absolute value of the magnetic field $|H_y|$ under antisymmetric illumination is shown. The color bar is in arbitrary units.}
\label{layers}
\end{figure}

It is known that the effective medium approach sometimes gives inaccurate description of nanostructured media. Significant spatial dispersion may arise in nanolayered materials~\cite{Avrutsky}, as well as in wire medium even in the long wavelength limit~\cite{Belov}.
Therefore it is important to justify the use of effective medium approximation in our study. We accomplish that by direct modeling the CPA eigenmode scattering by a layered nanostructure without use of effective parameters. The resulting field distribution is shown in Fig.~\ref{layers} for 12~nm unit cell thickness of upper and lower materials. The absorbing structure length yielding minimal scattering shifts from 770~nm to 800~nm due to the finite periodicity of the system.
Nevertheless, the field distribution resembles that of ideal homogenized metamaterial shown in Fig.~\ref{fig2}(b) and the total power scattered by the layered system is only 0.5\% of the incident power. The rest of the calculations will be performed for homogeneous metamaterials with effective permittivites given by Eqs.~(\ref{eqeff}).
\begin{figure}[!t]
\includegraphics[width=.8\columnwidth]{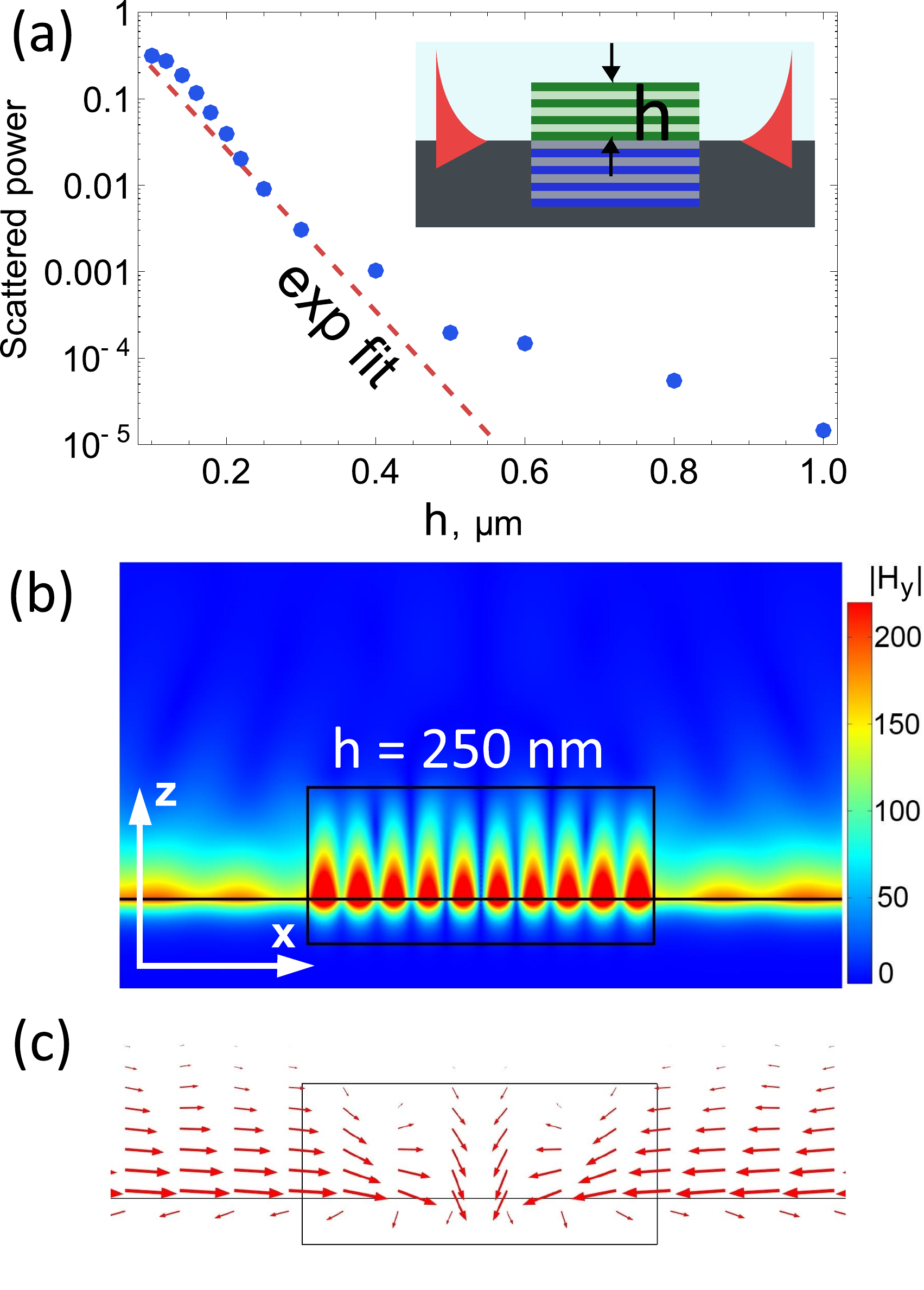}
\caption{(a) The total power scattered by the absorber as a function of the top anisotropic metamaterial thickness $h$. (b) Absolute value of the magnetic field $|H_y|$ created by coherent illumination of the system by the CPA eigenmode at $h=$~250 nm. The color bar is in arbitrary units. (c) The corresponding time-averaged Poynting vector map.}
\label{fig3}
\end{figure}

Now let us analyze the effect of the system geometry on robustness of coherent SPP absorption. Equation~(\ref{eq3}) defines scattering-free SPP transmission for an interface between two semi-infinite anisotropic materials. In reality, they would have finite thickness in $z-$direction, thus modifying SPP dispersion and field distribution. As a result, non-zero scattered power may appear both in plasmon and free space modes channels.

In order to illustrate the effect of finite absorber thickness, we numerically solve scattering of the CPA eigenmode by the finite system. The lower metamaterial thickness can be safely set to $h_{\rm low}=100$~nm without affecting the SPP mode as its field is localized in metal at sub-100 nm scale. The total power $W$ scattered from the SPP absorber (normalized by the incident power) is shown in Fig.~\ref{fig3}(a) as a function of the top metamaterial thickness $h$. 
Scattered power can be roughly estimated as the fraction of SPP energy which does not interact with absorbing metamaterial yielding an exponential fit $W \cong \exp \left( { - 2\operatorname{Im} k_z^{(1)}h} \right)$.
For thickness below 0.5~$\mu$m the dependence of scattered power nearly follows this exponential fit (discrepancy for larger values possibly stems from numerical errors in calculation of small variables).
The dependence also indicates that at $h=200$~nm the total scattered power is at the reasonable level of about 1\%.

The resulting magnetic field distribution and the Poynting vector map for the thickness of the upper metamaterial $h=250$~nm are presented in Figs.~\ref{fig3}(b) and (c). A slight modulation of the SPP field along the $x-$direction indicates that reflected SPP appears in a finite thickness system. The characteristic leaking pattern of electromagnetic field into free space can also be observed. Nevertheless, the Poynting vector map shown in Fig.~\ref{fig3}(c) confirms that most of the energy sinks into the anisotropic absorber without any significant energy fraction being leaked into free space.

\begin{figure}[!b]
\includegraphics[width=.8\columnwidth]{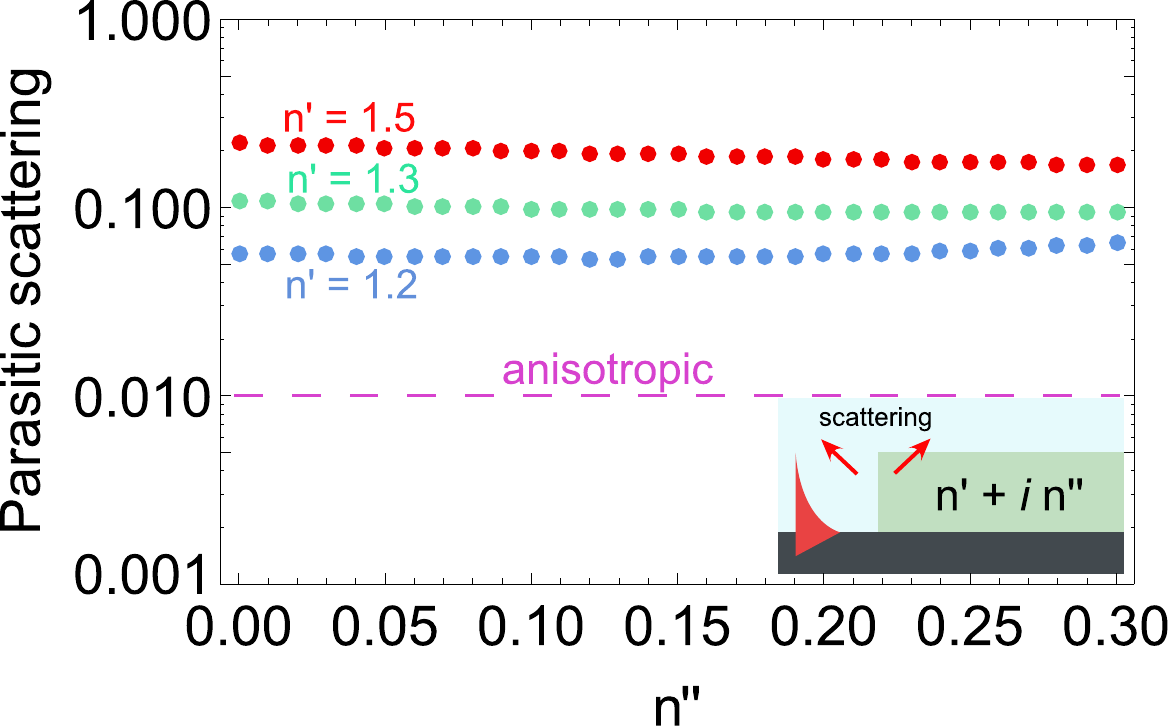}
\caption{Normalized free space scattering from an absorbing dielectric cover of 500 nm thickness lying on top of silver. Dashed line indicates the level of scattering achieved with matched anisotropic metamaterial for $h=250$~nm.}
\label{figCover}
\end{figure}

In order to highlight the advantage of our approach to coherent SPP absorption we demonstrate the performance of a similar absorbing system which does not exhibit the scattering-free property. The system is formed by a dielectric cover of height $h=500$~nm with refractive index $n=n'+i n''$ lying on top of semi-infinite silver. Such a cover may be realized as a dye doped polymer. The resulting dependence of normalized parasitic scattering from a single interface boundary on $n''$ is shown in Fig.~\ref{figCover} for a series of $n'$. As the results suggest, the amount of parasitic scattering in this approach is order of magnitude greater than with 250~nm height anisotropic structure. Such high free space scattering shows that the CPA regime can not be achieved with only an isotropic dielectric absorbing layer.

Finite absorber dimensions and spatial dispersion are not the only sources of parasitic SPP scattering. Realistic silver-air interface produces significant SPP scattering due to surface roughness. In order to evaluate the importance of surface roughness on parasitic scattering, we adopt the state of the art experimental data on propagation length of SPPs on silver films~\cite{Norris}. It was observed that surface roughness of about 0.8~nm decreases the propagation length on average by 5\%. In our case such attenuation is equivalent to loss of $\sim 0.25\%$ of energy upon SPP propagation over 1~$\mu$m. However, the recent advances in films technology enable fabrication of Ag films with smaller imperfections of 0.24~nm roughness (Ref.~\cite{Li}) which will result in even smaller scattering of the order of $0.1\%$.

Fabrication of the considered layered structure may be challenging. Nevertheless, we expect that it is possible with existing techniques and propose specific technological steps that can be applied. 
At first, Ag film is covered by a layer of photoresist (e.g. PMMA). Then, the focused ion beam technique is applied for fabrication of a slot for the absorbing structure. After that, the pulsed laser deposition method is used for metal/dielectric layers synthesis over the whole sample area. The desired absorbing structure is formed after lift-off process. This procedure is summarized in Fig. S1 of Supplementary Material.

The proposed concept of coherent absorber is not limited to MI interfaces, it is equally applicable to other SPP guided mode systems. The problem reduces to finding the parameters of a scattering-free interface, satisfying conditions~(\ref{eq3}) and yielding zero of one of the $S-$matrix eigenvalues. 
Similar design rules can also be applied to engineer a perfect absorption in different systems, including Dyakonov surface waves structures~\cite{Dyakonov}.

\section{Conclusion}
To conclude, we have proposed a design of a scattering-free coherent perfect absorber of surface plasmon polaritons. The absorber is represented by an interface between two uniaxial lossy materials, coupled on both sides to conventional metal-insulator interfaces. Parasitic scattering of incident SPPs into free space modes is eliminated by properly chosen permittivity tensors of the absorbing anisotropic materials. The feasibility of our approach is demonstrated via full-wave simulations with a spatially non-uniform metal-dielectric layered nanostructure. It is shown that the amount of scattered power can be accurately controlled by variation of the geometric dimensions of the nanostructure. We envision that our findings will have far reaching implications in the design of integrated nanophotonic systems, including plasmonic modulators and interference based plasmonic logical gates enabling flexible SPP manipulation.

\begin{acknowledgements}
We thank Dr. Dmitry A. Zuev and Dr. Alexander E. Krasnok for helpful discussion. The work was financially supported by the Advanced Research Foundation (Contract No 7/004/2013-2018) and RFBR project No 16-32-00444.
\end{acknowledgements}

\bibliography{plasmonCPA}

\end{document}